\documentclass[useAMS,usenatbib,onecolumn]{mn2e}
\usepackage{amsmath}
\usepackage{graphicx}
\usepackage{amssymb}
\usepackage{sidecap}
\usepackage{color}

\def\ba{\begin{eqnarray}}
\def\ea{\end{eqnarray}}
\def\be{\begin{equation}}
\def\ee{\end{equation}}

\def\eop{\mathcal{E}}
\def\bop{\mathcal{B}}
\def\vl{\vec{\ell}}

\def\be{\begin{equation}}
\def\ee{\end{equation}}

\newcommand{\ave}[1]{\left\langle#1\right\rangle}
\title{Lensing Simulation and Power Spectrum Estimation for High Resolution CMB Polarization Maps}

\author[Louis et al.]{
 Thibaut Louis$^1$\thanks{E-mail: Thibaut.Louis@astro.ox.ac.uk},  Sigurd N{\ae}ss$^1$, Sudeep Das$^2$, Joanna  Dunkley$^1$,  Blake Sherwin$^3$. \\
 $^1$     Sub-department of Astrophysics, University of Oxford, Keble Road, Oxford, OX1 3RH, UK. \\
 $^2$     Argonne National Laboratory, 9700 S.~Cass Ave., Lemont, IL 60439 \\
 $^3$    Joseph Henry Laboratories of Physics, Jadwin Hall, Princeton University, Princeton, NJ, USA 08544. }

\begin{document}


\label{firstpage}

\maketitle

\begin{abstract}
We present efficient algorithms for CMB lensing simulation and  power spectrum estimation for  flat-sky CMB polarization maps. We build a pure B-mode estimator to remedy  E to B leakage due to partial sky coverage. We show that  our estimators are unbiased, and consistent with the projected errors. We demonstrate our algorithm using simulated observations of small sky patches with realistic noise and weights for upcoming CMB polarization experiments. 
\end{abstract}


\section{Introduction}

The recent measurements of the cosmic microwave background (CMB) power spectrum by the Atacama Cosmology Telescope \citep[ACT,][]{Das:2013zf}, South Pole Telescope  \citep[SPT,][]{Story:2012wx}, and the {\it Planck} satellite \citep{Planck:2013kta} at small angular scales have provided important confirmation of the standard $\Lambda$CDM cosmological model, extending the measurements by the {\it WMAP} satellite \citep{Hinshaw:2003ex,Bennett:2012fp}, and earlier observations. 
The next generation of CMB experiments are focused on measuring the polarization of the CMB. The anisotropies are only $\sim$10\% polarized so an accurate measurement of the polarization power spectrum is challenging. A number of ground-based experiments are targeting this signal, with POLARBEAR  \citep{Kermish:2012eh}, SPTpol  \citep{Austermann:2012ga}, and ACTPol \citep{Niemack:2010wz} designed to measure  scales of a few arcminutes or less. 
These experiments aim to constrain the cosmological model by measuring the `E-mode' power spectrum that provides an independent probe of the scalar modes measured through the temperature fluctuations,
and by measuring the 'B-modes' generated due to the gravitational  lensing of E-modes by the dark matter distribution along the line-of-sight. These lensing B-modes are generated at small scales, and are more easily accessible to high resolution ground-based telescopes than the larger scale primordial B-modes directly sourced by gravitational waves.

In this paper we describe the estimation of E and B-mode power spectra from realistic observations of the CMB sky. The number of pixels for high resolution experiments is of order $\approx 10^{7}$, so a direct maximum likelihood method is computationally too expensive. Instead we rely on pseudo $C_{\ell}$ estimators  \citep{Bond:1998zw}. One of the main challenges in estimating the pseudo B-mode power spectrum arises since the E and B mode decomposition of the polarization field on a incomplete sky  induces leakage between the two modes. The discontinuity at the edges of the map mixes E and B modes, increasing the variance of the B-mode power spectrum. \citet{Smith:2005gi} and \citet{Smith:2006vq}  provide a  general solution to this problem by defining a pure B-mode power spectrum estimator that is not contaminated by this mixing. In Section 2, we adapt their algorithm for flat sky maps,  and demonstrate that the algebra simplifies considerably under the flat-sky approximation.

In Section 3, we introduce a novel technique for generating high resolution lensed CMB maps. Different methods have been proposed to do this, often using a remapping between pixels  \citep{Lewis:2005tp}  and an interpolation scheme. We present a hybrid method that combines pixel remapping and a Taylor  series decomposition of the lensed field. In Section 4, we use simulated observations from the ACTPol experiment to generate non-uniform realizations of the experimental  noise. We then test our lensing simulation and power spectrum estimation method, and its optimality, using Monte Carlo simulations. 

\section{E/B leakage in the flat sky approximation}

In this section we review the issue of leakage of power between polarization types due to incomplete sky coverage. This issue has been discussed in previous studies targeting observations over large areas \citep[e.g.,][]{lewis/etal:2002,Smith:2005gi,Smith:2006vq,grain/etal:2009,cao/fang:2009,zhao/baskaran:2010,bunn:2011,bowyer/etal:2011,grain/etal:2012}. We demonstrate how the `pure' estimators of the different polarization types can be applied in the flat-sky approximation, relevant to small patches of the sky.

\subsection{Notation}

For linear polarization, the Stokes parameter Q quantifies the polarization in the x-y 
direction and U quantifies it along axes rotated by $45^{\circ}$. 
Following e.g., \cite{born/wolf:1980}, the polarization tensor is given by
\ba
P_{ab}(\vec{x}) = \frac{1}{2}
\begin{pmatrix}
Q(\vec{x}) &  U(\vec{x}) \\
U(\vec{x}) & -Q(\vec{x})
\end{pmatrix}.
\ea
Any $2\times2$ symmetric traceless tensor can be uniquely decomposed into two parts of the form
$
\eop_{ab} A=(-\partial_{a}\partial_{b}+ \frac{1}{2}\delta_{ab} \nabla^{2})A
$
and 
$
\bop_{ab}B=\frac{1}{2} (\epsilon_{ac}\partial^{c}\partial_{b} + \epsilon_{bc}\partial^{c}\partial_{a})B
$
where A and B are scalar functions \citep[e.g.,][]{Kamionkowski:1996ks}. The Fourier modes $e^{i\vl \vec{x}}$ provide a basis for a scalar function in the plane, so one can define
\ba
(^{E}e^{i\vl . \vec{x}})_{ab} &=& N_{\vl} \eop_{ab} (e^{i\vl\vec{x}})= N_{\vl} \left( \ell_{a}\ell_{b}- \frac{\vl^{2}}{2}\delta_{ab} \right)e^{i\vl\vec{x}} \nonumber\\
(^{B}e^{i\vl . \vec{x}})_{ab} &=& M_{\vl}\bop_{ab} (e^{i\vl\vec{x}})=- \frac{M_{\vl}}{2} \left(\epsilon_{ac}\ell^{c}\ell_{b} + \epsilon_{bc}\ell^{c}\ell_{a}\right) e^{i\vl\vec{x}},
\ea
where $N$ and $M$ are normalization coefficients that satisfy the orthogonality relation
\ba
\int d^{2}x (^{E}e^{i\vl\vec{x}})^{*}_{ab} (^{E}e^{i\vl'\vec{x}})^ {ab} = \delta( \vl-\vl'),
\ea
and similarly for $^{B}e^{i\vl\vec{x}}$.
Expanding the polarization field in this basis, it can be expressed as a combination of parity even (E) and odd (B) modes
\ba
\label{eq:EB}
P_{ab}(\vec{x}) &=& \frac{1}{\sqrt{2}} \int d\vl E(\vl) (^{E}e^{i\vl\vec{x}})_{ab} +B({\vl}) (^{B}e^{i\vl\vec{x}})_{ab}, \nonumber\\
 E({\vl}) &=& \frac{2}{\ell^{2}} \int d^{2}x   P_{ab}(\vec{x})   \eop^{ab}(e^{-i\vl\vec{x}}), \nonumber\\
 B({\vl}) &=& \frac{2}{\ell^{2}} \int d^{2}x   P_{ab}(\vec{x})  \bop^{ab}(e^{-i\vl\vec{x}}). 
\ea
This is equivalent to the spin formalism introduced by \citet{Seljak:1996gy}. Defining $\phi_{\ell}$ as the angle between the vector $\vec{\ell}$ and the $\ell_{x}$ axis, E and B take the simple form
\ba
E({\vl}) &=& Q(\vl) \cos 2\phi_{\ell}+U(\vl)\sin 2\phi_{\ell},  \nonumber \\
B({\vl}) &=& -Q(\vl)\sin 2\phi_{\ell} +U(\vl) \cos 2\phi_{\ell}  
\label{eqn:eb}
\ea
or equivalently $E({\vl}) \pm i B({\vl})= e^{\mp 2i\phi_{\ell}}(Q(\vl) \pm i U(\vl))$.

\subsection{Partial sky coverage}

\begin{figure}
\begin{center}
\includegraphics[width=17cm]{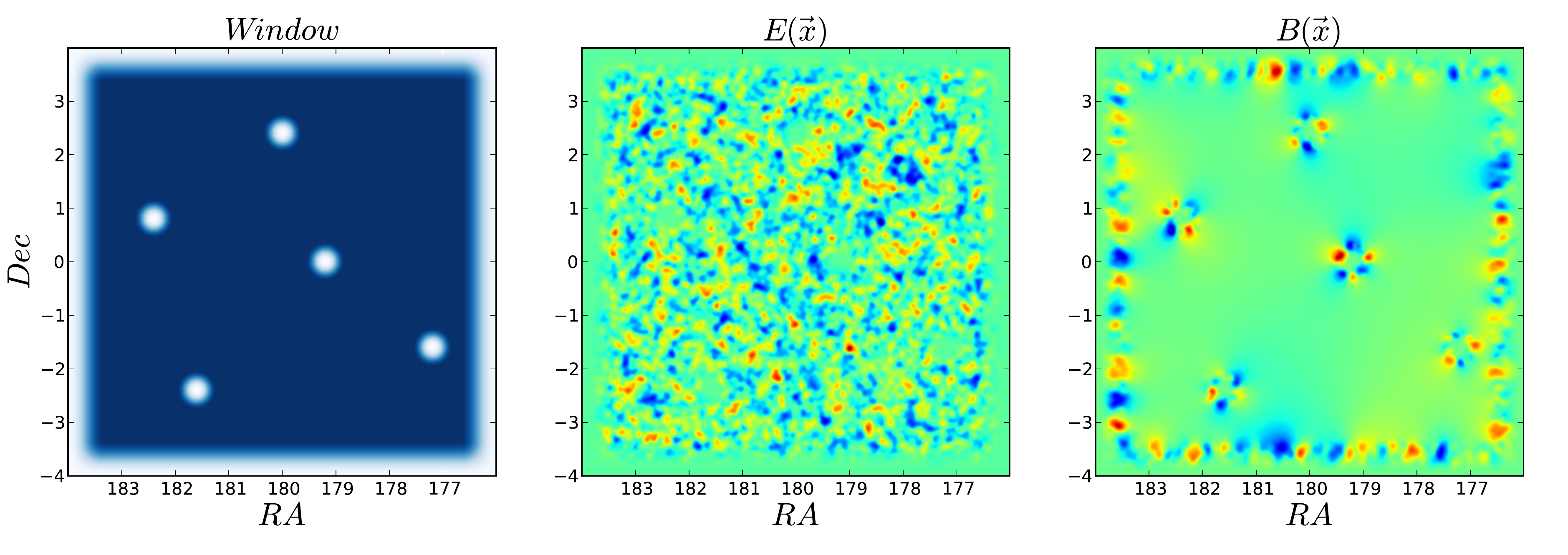}\\
\end{center}
\caption{{\bf Effect of sky cuts on the polarization pattern.}
A pure E-mode signal on the sky is observed through a window with a point source mask (left) leading to the estimated E-mode (centre) and B-mode  (right) maps. The leaked E-modes show up as spurious signal in the B-mode map localized around the discontinuities of the window function. 
}
\label{Fig:Window}
\end{figure}

Observations with modern high resolution experiments are typically performed on a small fraction of the sky, of area $\Omega$ \citep[e.g.,][]{Story:2012wx,Das:2013zf}. The observed region can be described by a window function
\ba
W(\vec{x}) &=& \left\{ \begin{array}{rcl} w( \vec{x}) & \mbox{if} &  \vec{x} \in \Omega \\ 0 & &\mbox{otherwise}
\end{array}\right. 
\ea
which modifies the observed Q and U components such that $\tilde{Q}(\vec{x})=W(\vec{x})Q(\vec{x})$ and $\tilde{U}(\vec{x})=W(\vec{x})U(\vec{x})$.
Propagating the effects of the window function into the $\tilde{E}$ and $\tilde{B}$ modes calculated as in Eqn.~\ref{eqn:eb}, the window functions mix E and B modes to give the modified modes
\ba
\label{eq:NonPureEstimator}
\tilde{E}({\vl})&=&  \int d\vec{\ell'} W(\vec{\ell}-\vec{\ell'})  [  E({\vl'}) \cos  2(\phi_{\ell'} - \phi_{\ell})  - B({\vl'}) \sin  2(\phi_{\ell'} - \phi_{\ell})  ],   \nonumber \nonumber \\
\tilde{B}({\vl}) &=&  \int d\vec{\ell'} W(\vec{\ell}-\vec{\ell'})  [  E({\vl'}) \sin  2(\phi_{\ell'} - \phi_{\ell})  + B({\vl'}) \cos  2(\phi_{\ell'} - \phi_{\ell})  ].  
\ea
The power spectra of these modified modes are then
\ba
\langle \tilde{E}({\vl}) \tilde{E}^{*}({\vl}) \rangle=   \int d\vec{\ell'} | W(\vec{\ell}-\vec{\ell'})|^{2}  [  C_{EE}(\ell') \cos^{2}  2(\phi_{\ell'} - \phi_{\ell})  + C_{BB}(\ell')\sin^{2}  2(\phi_{\ell'} - \phi_{\ell})  ],   \nonumber \\
\langle \tilde{B}({\vl}) \tilde{B}^{*}({\vl}) \rangle=   \int d\vec{\ell'} | W(\vec{\ell}-\vec{\ell'})|^{2}  [  C_{EE}(\ell') \sin^{2}  2(\phi_{\ell'} - \phi_{\ell})  + C_{BB}(\ell')\cos^{2}  2(\phi_{\ell'} - \phi_{\ell})  ].
\ea
Since $C_{EE}$ is expected to be an order of magnitude larger than $C_{BB}$, this mixing therefore leads to significant contamination of the B-modes by the leaked-in E-modes. This is illustrated in Figure \ref{Fig:Window}, which also highlights that the leakage  is localized close to the edges of the map, or to the edges of holes due to masking of bright point sources. This leakage will increase the variance of the measured $C_{BB}$, even when an unbiased estimator is constructed for $ C_{BB}$ and $ C_{EE}$ by inverting the convolutions.  

\subsection{Pure estimators}

A solution to the problem of E/B mixing has been proposed by \citet{Smith:2005gi,Smith:2006vq}. Rather than deconvolving $E(\ell)$ and $B(\ell)$, the window function can be included directly in the projection operator (Eqn.~\ref{eq:EB}), such that
\ba
E^{pure}({\vl})&=& \frac{2}{\ell^{2}} \int d^{2}x   P ^{ab}(\vec{x})  \eop_{ab}( W(\vec{x}) e^{-i\vl\vec{x}}) \\
B^{pure}({\vl})&=& \frac{2}{\ell^{2}} \int d^{2}x   P ^{ab}(\vec{x})  \bop_{ab}( W(\vec{x}) e^{-i\vl\vec{x}}). 
\ea
Using this method, any ambiguous modes are projected out and pure E and B modes are recovered. The window function and its first derivative must be zero at the edges of the map to avoid generating spurious B modes.

Here we show how this method can be simply applied in the flat sky approximation. By applying the product rule to the differential projection operators on the right hand sides of the above equations, we obtain expressions for the pure E and B modes. The $B^{pure}({\vl})$ mode is given by
\ba
\label{eq:pureB}
B^{pure}({\vl})  &=& \frac{2}{\ell^{2}} \int d^{2}x  W(\vec{x})  P ^{ab}(\vec{x})  \bop_{ab}( e^{-i\vl\vec{x}}) +  \frac{1}{\ell^{2}} \int d^{2}x   \left[  2 Q \partial_{x}  \partial_{y} W+ U ( \partial_{y}^{2}- \partial_{x}^{2}) W \right] e^{-i\vl\vec{x}}  \nonumber \\
&-& \frac{2i}{\ell} \int d^{2}x  \left[ Q(\vec{x}) ( \partial_{y} W   \cos\phi_{\ell}  +  \partial_{x} W  \sin \phi_{\ell}  ) + U(\vec{x}) (  \partial_{y} W  \sin \phi_{\ell} -  \partial_{x}W \cos \phi_{\ell}) \right] e^{-i\vl\vec{x}}.
\ea
The first term is the standard ``naive'' B mode estimator and the second and third terms cancel the window-induced leakage from E to B modes, involving derivatives of the window function.
This expression is convenient for numerical uses as it does not require the calculation of derivatives of noisy data. 
The pure estimator removes the E/B leakage, but the remaining mode coupling effect induced by applying a window to the observed sky still needs to be deconvolved. 

To work out this effect, we can simplify the algebra if we express the pure estimator in term of the $\chi$ variables \citep[e.g.,][]{lewis/etal:2002}, with
\ba
\chi_{E}(\vec{x}) &=& - \frac{1}{2}  \left[ \bar{\eth}\bar{\eth}(Q+iU)(\vec{x}) +\eth \eth(Q-iU)(\vec{x})  \right],  \nonumber \\ 
\chi_{B}(\vec{x}) &=& \frac{i}{2}  \left[ \bar{\eth}\bar{\eth}(Q+iU)(\vec{x}) -\eth \eth(Q-iU)(\vec{x})  \right], 
\ea
where the spin raising and spin lowering operators are defined as
$\eth = - (\partial_{x}+i \partial_{y})$, and  $\bar{\eth} = - (\partial_{x}-i \partial_{y})$.
After some algebra we find simple expressions for the two pure modes,\footnote{This is most directly verified by inserting the definition of $\chi_{B/E}$ into the above integrals and integrating by parts twice; the result obtained is identical to the original form of the pure estimators.}
\ba
B^{pure}({\vl})  &=& \frac{1}{\ell^{2}} \int d\vec{x} \chi_{B}(\vec{x}) W(\vec{x}) e^{-i\vl \vec{x}}, \nonumber \\ 
E^{pure}({\vl})  &=& \frac{1}{\ell^{2}} \int d\vec{x} \chi_{E}(\vec{x}) W(\vec{x}) e^{-i\vl \vec{x}}.
\label{eqn:pure_eb}
\ea
The simple expressions for the pure estimators using the $\chi$ variables are convenient for computing this mode-to-mode coupling.
Noting that
\ba
\chi_{E} (\vec{x})  &=& \int d\vl E(\vl) \ell^{2} e^{i\vl \vec{x}},   \nonumber \\ 
\chi_{B} (\vec{x})  &=& \int d\vl   B(\vl) \ell^{2} e^{i\vl \vec{x}}, 
\ea
we find that the coupling between modes is
\ba
 \langle B^{pure}({\vl}) B^{*pure}({\vl}) \rangle&=&\frac{1}{\ell^{4}}  \left \langle  \int d\vec{x} \chi_{B}(\vec{x}) W(\vec{x}) e^{-i\vl \vec{x}} \int d\vec{x'}   \chi_{B}(\vec{x'}) W(\vec{x'}) e^{i\vl \vec{x'}} \right \rangle, \nonumber \\ 
 &=& \frac{1}{\ell^{4}}  \int d\vl' |W(\vl-\vl')|^{2} \ell^{'4} C_{BB} (\vl')
\label{eqn:mode_eb}
\ea
for the B-modes, and
\ba
 \langle E^{pure}({\vl}) E^{*pure}({\vl}) \rangle&=& \frac{1}{\ell^{4}}  \int d\vl' |W(\vl-\vl')|^{2} \ell^{'4} C_{EE} (\vl')
\ea
for the E-modes. Here we have used $\langle  \chi_{B}(\vl') \chi^{*}_{B}(\vl'') \rangle=\delta( \vl'-\vl'') \ell^{'4} C_{BB} (\vl') $. 
In practice using the pure E mode power spectrum estimator is unnecessary since the B-to-E leakage is small so the advantage of using a pure estimator is lost and results in a loss of sensitivity. We choose to use a hybrid approach  \citep{grain/etal:2012}, where the B mode power spectrum is computed using the pure formalism (Eqn.~\ref{eq:pureB}) and the E modes power spectrum is computed via the standard pseudo power spectrum formalism.

\section{Generating gravitationally lensed simulations}

\begin{figure}
\begin{center}
\includegraphics[width=13cm]{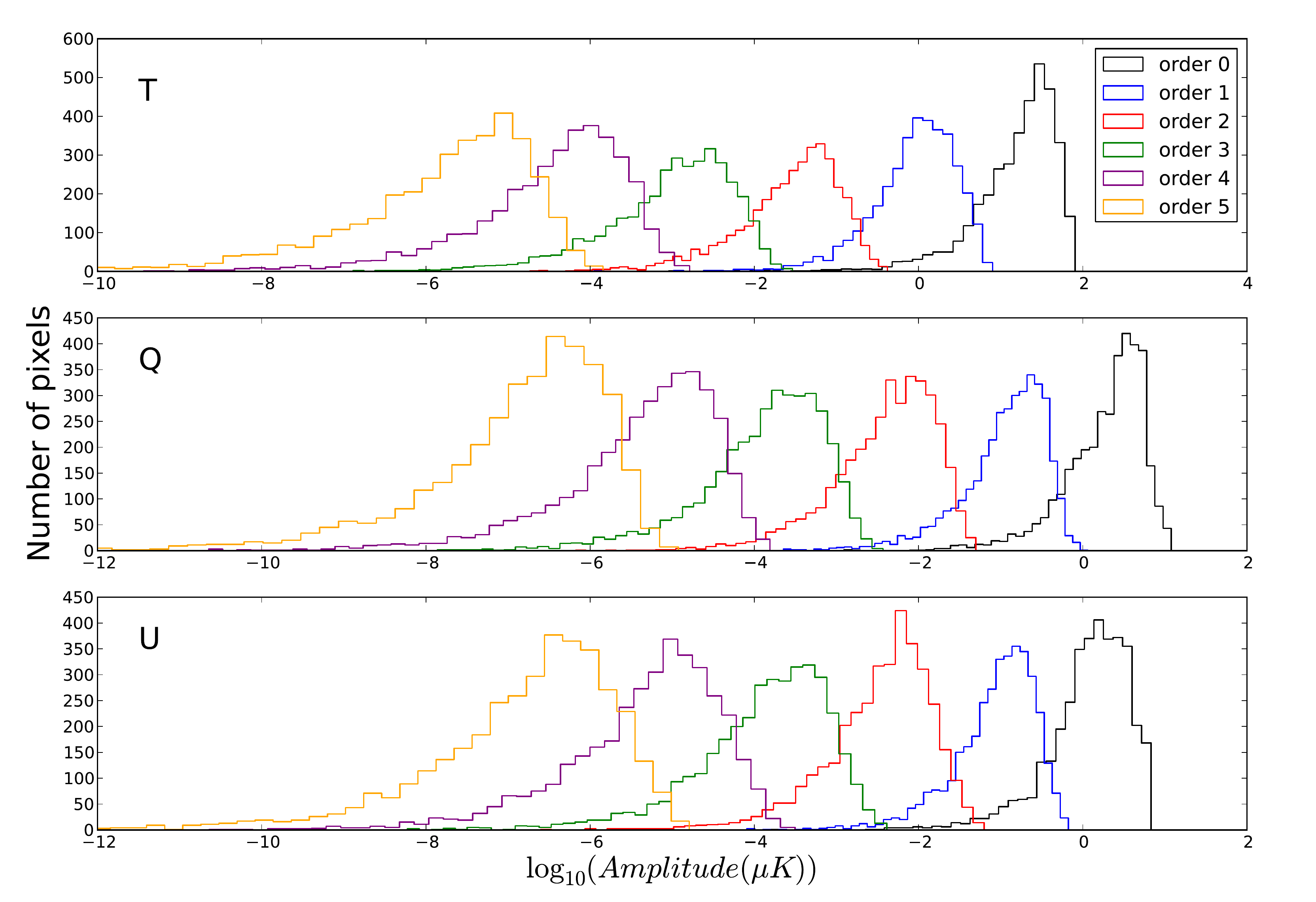}\\
\end{center}
\caption{{\bf Convergence of the Taylor series in pixel space}
We represent the contribution of each higher order term of the Taylor series by showing the histogram of its pixel distribution. The convergence of the series is fast, each term being $\approx 60$ times smaller than the preceding one. The contribution of the third order term is  of order $10^{-1} \mu$K for T and $10^{-2} \mu$K for Q and U. }
\label{Fig:Histo}
\end{figure}

On their way from the surface of last scattering, the photons are deflected
by the gravitational field of the intervening large scale structure. Accurate lensing simulations are essential for recovering the statistical property of the observed CMB. In the
weak field limit, the lensing results in a simple remapping of the temperature by
a deflection angle $\vec{\alpha} = \nabla \phi$, where $\phi$ is the lensing
potential, a line of sight integral over the matter distribution. The same applies to the
polarisation field:
\ba
\tilde{T} (\vec{x}) &=& T( \vec{x}+ \nabla  \phi)  \nonumber \\
\tilde{P}_{ab} (\vec{x}) &=&  P_{ab}( \vec{x}+ \nabla \phi).
\ea
This conceptually simple remapping is complicated by the fact that we
cannot work directly with the continuous real-space map, only with
discrete pixelizations of it. Three main approaches have been suggested
for implementing this remapping in simulations  \citep{Lewis:2005tp}:
\begin{enumerate}
	\item {\bf Go directly from frequency domain to the lensed positions}, i.e.
		$T(\vec x + \vec \alpha) = \sum_{\vec\ell} \mathcal{F}^{-1}_{\vec\ell}(\vec x + \vec \alpha) T_{\vec\ell}$,
		where $\mathcal{F}^{-1}_{\vec\ell}(\vec x)$ is the inverse Fourier transform operator,
		and $T_{\vec\ell}$ are the harmonic coefficients of the unlensed map. This approach
		is exact, but computationally inefficient because the shifted positions will not
		in general form a regular grid, and one hence cannot use Fast Fourier Transforms (FFT).
	\item {\bf Taylor expand the field}:
		This is straightforward, but has been found
		to converge slowly at small scales.
	\item {\bf Pixel remapping}: Truncate the displacement to the nearest pixel, and read off the
		corresponding pixel value. This remapping must be done at much higher resolution than
		the physical scales of interest in the map in order to avoid pixelization errors, and
		hence comes at a large cost both in terms of CPU-time and memory. It is therefore
		sometimes combined with pixel-space interpolation schemes.
\end{enumerate}
We present a simple modification to the Taylor expansion method that addresses its
slow convergence. In general, the Taylor expansion of a function $f(x)$ around a point $x_0$
becomes less accurate as the distance from $x_0$ grows, and conversely, the expansion can
be truncated earlier if one can expand around a point close to where one wishes to evaluate
the function.
The Taylor remapping method above expands $T(\vec x + \vec\alpha)$ around the point
$\vec\alpha=0$, and the reason for the slow convergence is that $\vec\alpha$ can be relatively
large. A better choice is to expand around the pixel center $\vec \alpha_0$ closest to $\vec \alpha$, which
is already exactly available, resulting in the following expansion
\ba
\tilde{T}(\vec{x})=T(\vec{x}+\vec{\alpha}_{0}+ \vec{\Delta \alpha})&=& T(\vec{x}+\vec{\alpha}_{0})+\Delta \alpha^{c} [\partial_{c} T ]({\vec{x}+\vec{\alpha}_{0}}) + \frac{1}{2}\Delta \alpha^{c} \Delta \alpha^{d}  [\partial_{c}\partial_{d} T ]({\vec{x}+\vec{\alpha}_{0}})  + ... 
\ea
The derivatives can be computed in Fourier space
\ba
\tilde{T}(\vec{x})&=& T(\vec{x}+\vec{\alpha}_{0}) + \sum_{n=1}^{\infty} \frac{i^{n}}{n!}  \left[ \int (\Delta \alpha^{x}  \ell_{x} +  \Delta \alpha^{y}  \ell_{y})^{n} T(\vl) e^{i \vl \vec{x}} d\vl  \right] _{\vec{x}+\vec{\alpha}_{0}} 
\ea
In practice we truncate the expansion at order N
\ba
\tilde{T}_{N}(\vec{x})&=&  \sum^{N}_{n=0} \sum_{ k \le n}  \frac{(\Delta\alpha^{x})^{n-k} (\Delta\alpha^{y})^{k}}{k!(n-k)!}  \mathcal{F}^{-1}(i^n \ell_{x}^{n-k} \ell_{y}^{k} \mathcal{F} T)(\vec x + \vec\alpha_0) .
\ea
Here we use the FFT to compute each of the derivatives.
\begin{figure}
\begin{center}
\includegraphics[width=13cm]{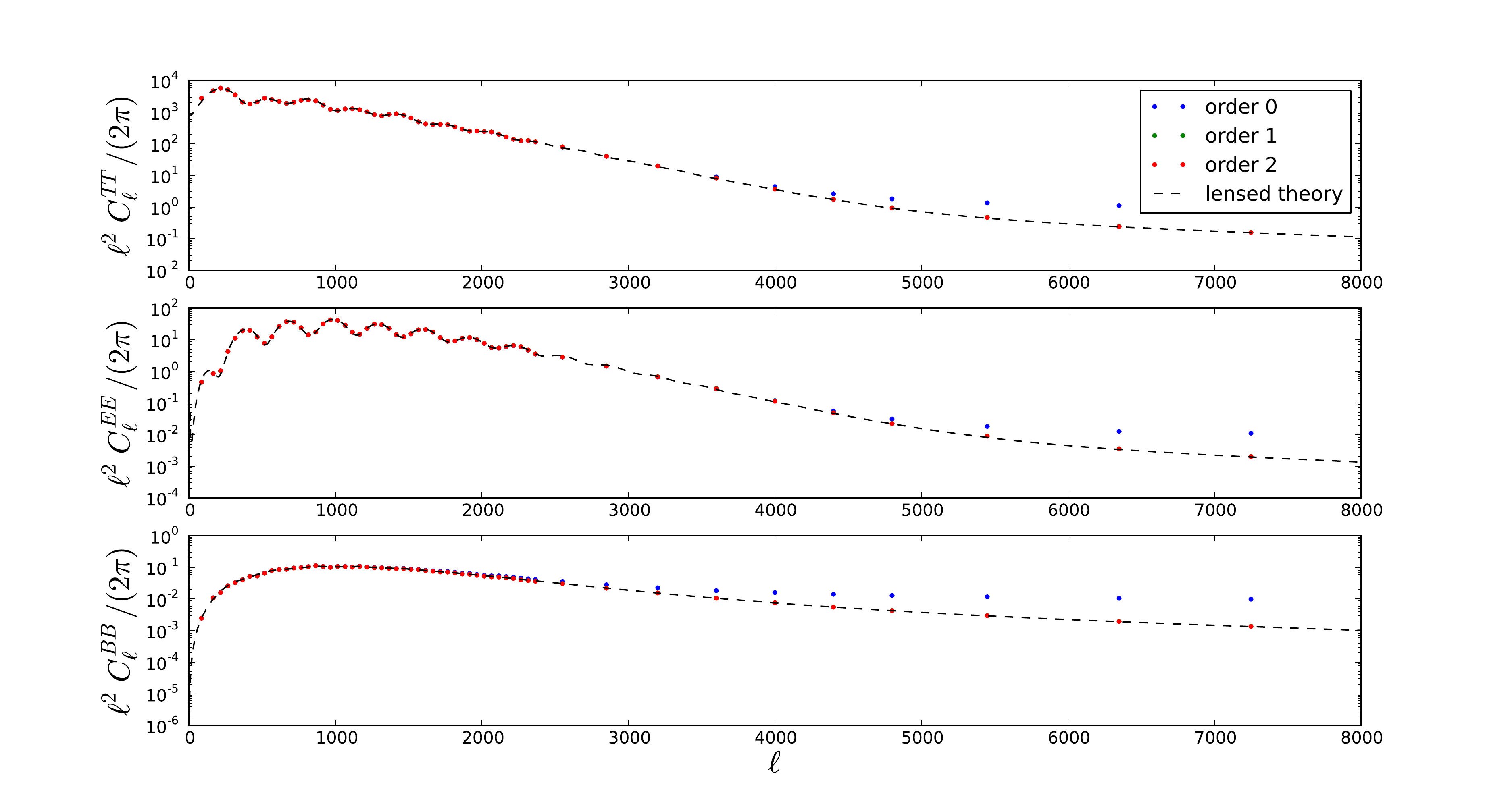}\\
\end{center}
\caption{{\bf Convergence of the Taylor series: power spectra }
We compute the temperature (TT) and polarization (EE, BB) power spectra of the series truncated at different orders. Convergence is achieved by second order in the expansion.
}
\label{Fig:PowerSpectra}
\end{figure}
Expanding around $\vec\alpha_0$ ensures that the Taylor expansion will at most need to
extrapolate by half a pixel in any direction, which ensures that all scales present in the
input map will converge rapidly.
This is effectively a hybrid between the pixel remapping and Taylor expansion methods,
but unlike normal pixel remapping one does not need to work at higher resolution than the
map that is being lensed.
Each term in this expansion can be computed at the cost of $(n+1)$ FFT. 
When expanding around $\vec\alpha_0$, we find that the series converges rapidly. In Figure~\ref{Fig:Histo} we show the pixel histograms of a part of the maps for each of the first
6 terms in the expansion. We find that the contribution falls by a factor of $\sim 60$
for each order for a $13^\circ\times13^\circ$ lensed noiseless CMB simulation with
0.5' pixel size. 

We also computed the bias for each order using 60
such simulations, shown in Figure~\ref{Fig:PowerSpectra}, and found that truncating the series
at second order is an excellent approximation for any realistic CMB experiment.
For comparison, the old method of expanding around $\vec\alpha=0$ requires more than
20 orders to converge at this resolution, which given the quadratic scaling
of the method corresponds to a performance difference of a factor of $\sim 50$.
Using this method, lensing a $13^\circ\times13^\circ$ patch of the sky at 0.5'
resolution takes only a few seconds on one processor, and only requires a few times
the memory that a single map takes up.
While the method is formulated in the context of the flat sky in this case, it also generalizes trivially to
the full, curved sky.

\section{Implementation on realistic observations}
\label{sec:observ}

High resolution ground-based experiments are observing small patches of the sky, measuring both the temperature and polarization of the CMB. There are also a set of lower resolution experiments underway targeting larger regions of sky in order to constrain or measure gravitational waves, but these will require analysis on the curved sky. In this section we test our power spectrum estimation method on simulated data, using a specific example of a subset of observations expected from the ACTPol experiment where the flat-sky approximation is appropriate.

\begin{figure}
\begin{center}
\includegraphics[width=19cm]{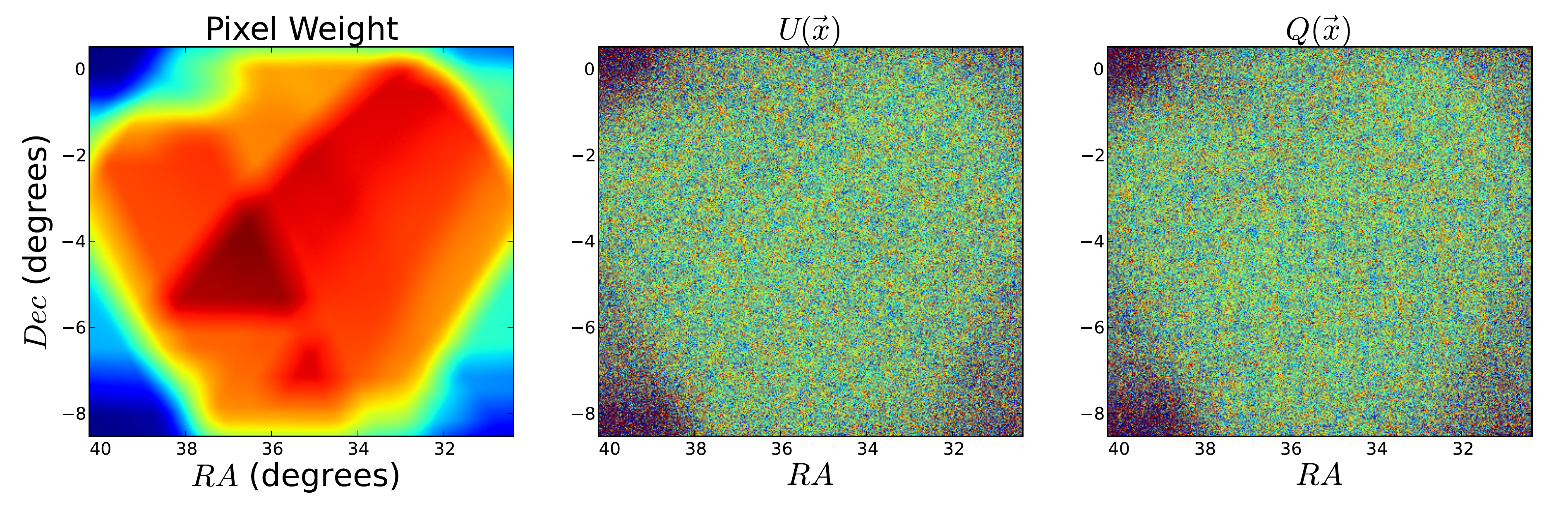}\\
\end{center}
\caption{{\bf Realization of the noise}, for a U and Q map (centre and right)  generated using a simulated pixel weight map (left). This represents the number of observations per pixel for an inhomogeneous survey, and is taken from a simulation for the ACTPol experiment.}
\label{Fig:InhomogeneousNoise}
\end{figure}

Here we assume that 4 patches of the sky, for a total area of $\sim 300$ deg$^2$, are observed to a noise level of $5.7~\mu$K/arcmin in temperature. The expected coverage of a patch is non-uniform due to the scanning strategy of the telescope; the expected statistical weight associated with each pixel is shown in the left panel of Figure \ref{Fig:InhomogeneousNoise} for one of the patches. ACTPol will also target a larger region of the sky. For comparison, the PolarBear experiment is targeting three $225$ deg$^2$ regions at $6~\mu$K/arcmin sensitivity in temperature \citep{Kermish:2012eh}, and SPTPol have initially targeted a 100 deg$^2$ region, with the goal to cover 625 deg$^2$ to $5~\mu$K/arcmin \citep{Austermann:2012ga}.

\subsection{Estimated power spectra}

We simulate data for each patch of sky in four subsets, generating independent maps of identical coverage and equal depth as in \citet{Das:2010ga,Das:2013zf}. We refer to each subset as a `split'.  
We convolve the simulation with a spherically symmetric gaussian beam with FWHM of $1'$,  and we simulate an inhomogeneous noise realization by convolving the weight map for each patch with a  $5.7~\mu$K/arcmin noise realization. 
We `prewhiten' the temperature maps as defined in \citet{Das/etal:2009}. Here, the maps are convolved in real space with kernels designed to make the power spectrum as flat as possible, to reduce aliasing of power due to the point source mask.
We then apply a 5' (apodized with a $0.3^\circ$ cosine kernel) point source mask  to account for the possible contamination from polarized extragalactic point sources.

We compute the binned cross-power spectrum $C^{iX\times jY}_{b}$ between maps $i$ and $j$, for polarization types $X$ and $Y$, using the pure estimators for B (Eqn.~\ref{eq:pureB}) and a standard Fourier transform for T and E as discussed in section 2. The estimated  spectrum is then given by
\be
\widetilde C_b^{iX\times jY}  = \sum_{b'} M^{XY}_{bb'} C^{iX\times jY}_{b'},
\ee
where the mode coupling matrix is 
\be
M^{XY}_{bb'} = \sum_{\vl,\vl'} P_{b\vl} |W^{XY}(\vl-\vl')|^{2} \left(\frac{\ell'}{\ell} \right)^{\beta_{XY}} ( F_{\ell'}^{XY})^2 Q_{\vl' b'}. 
\ee

Here $\beta_{XY}=2[ \delta_{BX}+   \delta_{BY}]$, i.e., for the pure-mode BB spectrum $\beta=4$, but $\beta=0$ for TT and EE.  
The window function $W^{XY}(\vl)$ is  a product of the point source mask, the $n_{\rm obs}$ weight map, and a $0.7^\circ$ cosine apodization at the edges (\citet{Smith:2006vq}), with a geometrical correction for the E modes (Eqn.~\ref{eq:NonPureEstimator}). The function $F_{\ell}^{XY}$ is the product of the beam, a pixel window function and the transfer function of the prewhitener for the temperature power spectrum.
Here $P_{b\vl}$ is a binning matrix, and $Q_{\vl b}$  is  an interpolation matrix, the binning being defined as a set of annuli in the 2 dimensional power spectrum space. Here we choose a minimal bin size of $\Delta \ell = 100$. 
Each of the mode coupling matrices is computed exactly and inverted in order to recover an unbiased spectrum.

\begin{figure}
\begin{center}
\includegraphics[width=19cm]{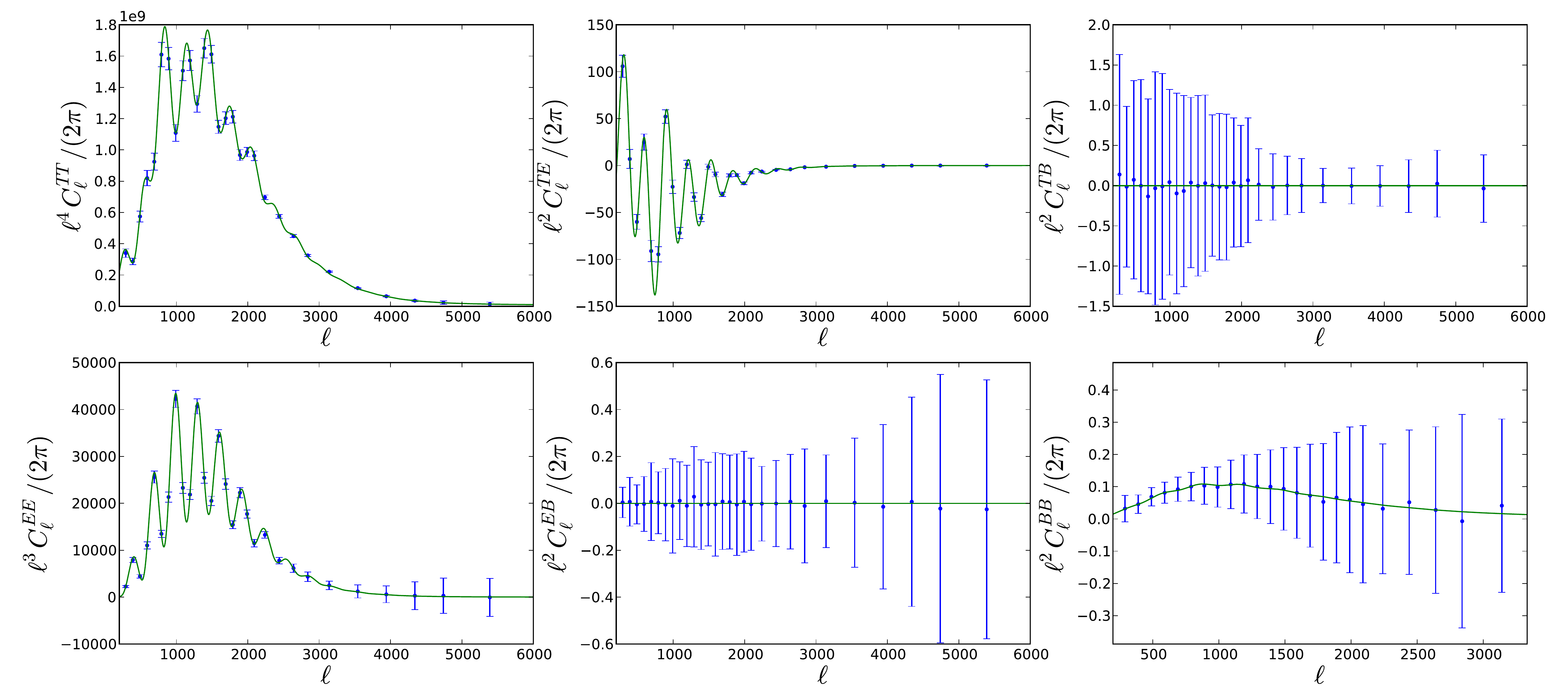}\\
\end{center}
\caption{{\bf Power spectra estimated from temperature and polarization maps.} This shows the average binned spectra estimated from 720 Monte Carlo simulations, with errors estimated from the $1\sigma$ dispersion. The B-mode spectra are derived using the pure estimator, to avoid leakage from the E-mode spectrum. }
\label{Fig:PS}
\end{figure}

We compute the spectra for 720 realizations of the noise and CMB, each with a different realization of the gravitational lensing potential. The cosmological model we use is the best-fitting $\Lambda$CDM model with no tensor contribution, so the B-mode signal comes only from gravitational lensing. The mean recovered spectra for TT, EE, BB, and the cross-correlation spectra are shown in Figure \ref{Fig:PS}, together with the estimated $1\sigma$ error bar for a single realization derived from the scatter of the simulations. The recovered power spectra are consistent with the input power spectra at the 0.1 $\sigma$ level in the interval $500 < \ell < 6000$.

\subsection{Power spectrum uncertainties}

\begin{figure}
\begin{center}
\includegraphics[width=19cm]{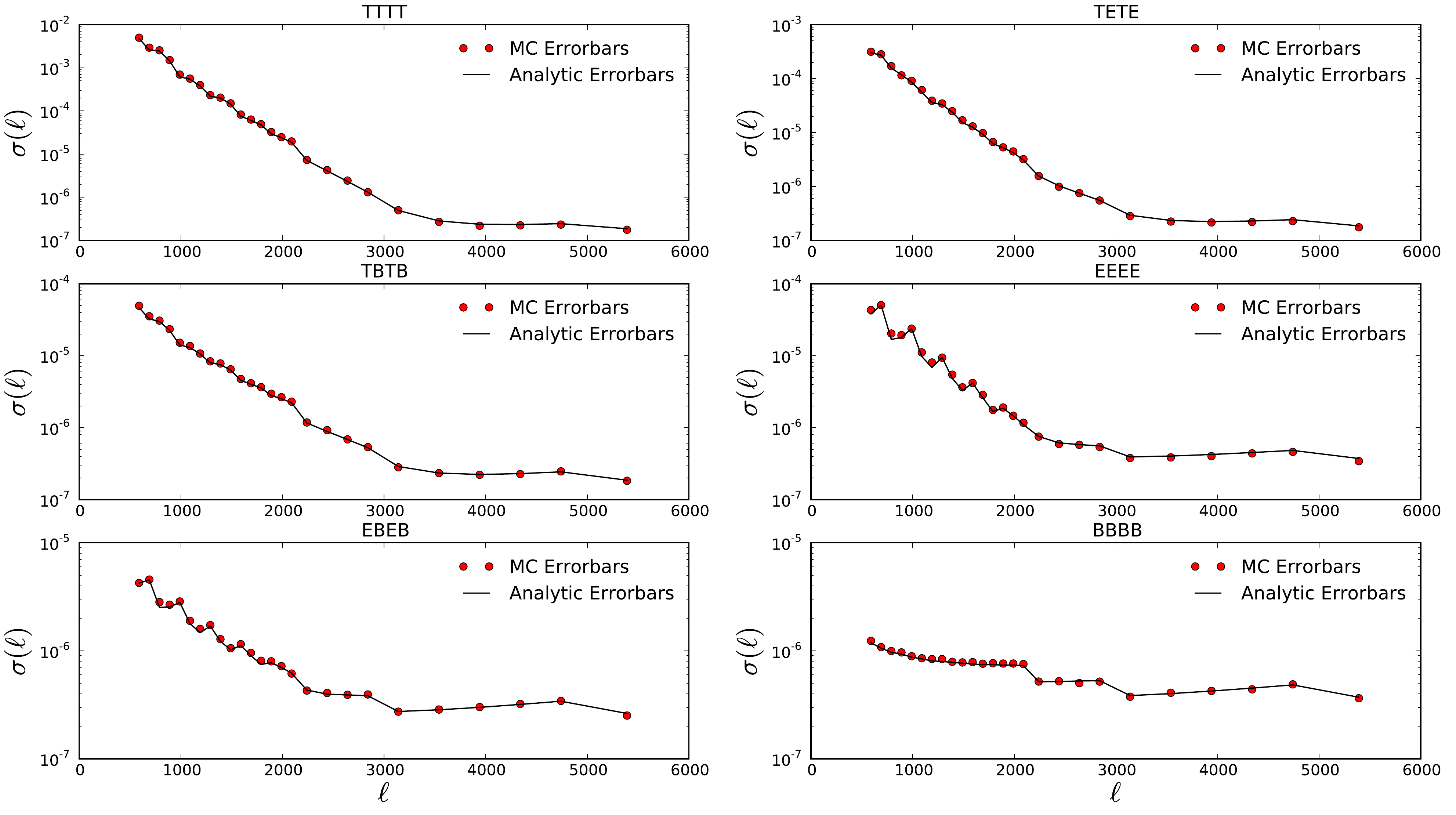}\\
\end{center}
\caption{{\bf Comparison between Monte Carlo scatter and analytic errors for each cross spectrum for one of the patches.}  They agree at the 15 per cent level for  $500 < \ell < 6000$, indicating that all sources of leakage are subdominant for these modes, The analytic estimate does not include the non-Gaussian contribution from lensing, but the noise in our simulation is high enough for this effect to be subdominant.
}
\label{Fig:Error}
\end{figure}

Using the Monte Carlo simulations, we can compare the errors derived from the internal scatter, with an analytic estimate. This provides a measure of the optimality of this method.  The analytic covariance in a single bin assuming no leakage is given by
\ba
 \Theta^{(X \times Y);( W \times Z)}_{bb}&=&\frac{1}{\nu_b}\left( C_{b} ^{X \times W}C_{b} ^{Y \times Z} +C_{b} ^{X \times Z}C_{b} ^{Y \times W} \right)  \nonumber \\
 &+& \frac{1}{\nu_b n_{d}} \left( C_{b} ^{X \times W} N_{b} ^{Y \times Z} +C_{b} ^{Y \times Z}  N_{b} ^{X \times W}+C_{b} ^{X \times Z} N_{b} ^{Y \times W}+C_{b} ^{Y \times W} N_{b} ^{X \times Z} \right) \nonumber \\
 &+& \frac{1}{\nu_b n_{d}(n_{d}-1)}  \left( N_{b} ^{X \times W} N_{b} ^{Y \times Z} +N_{b} ^{X \times Z}N_{b} ^{Y \times W}  \right),
\ea
where $n_{d}$ is the number of splits and $\nu_b$ is the number of modes per bin, corrected for the effect of the window function. $C^{Y \times Z}_{b}$ is the theoretical power spectrum, and $N^{Y \times Z}_{b}$ is the noise power spectrum, given by  $C^{Y \times Z}_{b,{\rm auto}}-C^{Y \times Z}_{b,{\rm cross}}$. The derivation of this expression is given in the Appendix. This does not include the non-Gaussian part of the covariance due to the effect of lensing, described in \citet{BenoitLevy:2012va}. This is a subdominant part of the error for the noise levels we consider here, but introduces correlations between bins. We find that the analytic error bars agree with the $1\sigma$ dispersion from the simulations at the 15\% level for $500 < \ell < 6000$, as shown in Figure \ref{Fig:Error}, indicating that all sources of leakage on these scales are subdominant. The error is dominated by cosmic variance at large scales, but at smaller scales is noise dominated. This agreement is promising and demonstrates the power of the pure estimator to recover the B-mode spectrum.
In practice these spectra will be used to construct a likelihood for testing cosmological models, such that 
\be
-2 \ln L = ({\tilde C}_b-C^{\rm th}_b)^TQ^{-1}({\tilde C}_b-C^{\rm th}_b).
\ee
The full covariance matrix, $Q$, can be estimated numerically from the simulations, or analytically, and the binned theory spectra $C^{\rm th}_b$ computed using bandpower window functions. The realistic likelihood will also include the lensing deflection spectrum, estimated from higher point statistics of the map, and appropriate cross-correlations.

\section{Conclusions}

A number of issues arise in the analysis of high resolution CMB polarization maps, one of the most significant being the leakage of E-mode  into B-mode polarization due to observing a limited region of sky. In this paper we have described a simple method for estimating the power spectrum in the flat sky approximation that minimizes this leakage. It draws on an existing all-sky method using a `pure' estimator for the B-mode,  and simplifies the approach for the flat sky. This will be appropriate for small regions observed by current CMB experiments including ACTPol, SPTPol, and PolarBear. Using a suite of Monte Carlo simulations with realistic noise levels for upcoming experiments, we have demonstrated our ability to recover unbiased and quasi-optimal power spectra. 

To test the robustness of any power spectrum method requires accurate simulations. The B-mode polarization spectrum at small angular scales is sourced solely from the gravitational lensing of the E-mode signal. We have shown how high resolution lensed CMB maps can be rapidly and accurately simulated using a hybrid approach between pixel remapping and interpolation in harmonic space. This method, which has advantages over the standard pixel-space interpolation approach, also has the potential to be extended to full sky spherical maps.

\section*{acknowledgments}
We thank David Spergel and Jeff McMahon for useful discussions and Mike Nolta for providing simulated coverage for the upcoming ACTPol experiment. SD acknowledges support from the David Schramm Fellowship at Argonne
National Laboratory and the Berkeley Center for Cosmological Physics fellowship. Funding from ERC grant 259505 supports JD, TL and SN.

\bibliography{act,PSbib}

\begin{thebibliography}{25}
\expandafter\ifx\csname natexlab\endcsname\relax\def\natexlab#1{#1}\fi

\bibitem[{Ade {et~al}\mbox{.}(2013)Ade {et~al.}}]{Planck:2013kta}
Ade P., {et~al.}, 2013, arXiv:1303.5075

\bibitem[{Austermann {et~al}\mbox{.}(2012)Austermann, Aird, Beall, Becker,
  Bender, {et~al.}}]{Austermann:2012ga}
Austermann J., Aird K., Beall J., Becker D., Bender A., {et~al.}, 2012,
  Proc.SPIE Int.Soc.Opt.Eng., 8452, 84520E

\bibitem[{Bennett {et~al}\mbox{.}(2012)Bennett, Larson, Weiland, Jarosik,
  Hinshaw, {et~al.}}]{Bennett:2012fp}
Bennett C., Larson D., Weiland J., Jarosik N., Hinshaw G., {et~al.}, 2012

\bibitem[{Benoit-Levy, Smith \& Hu(2012)Benoit-Levy, Smith, \&
  Hu}]{BenoitLevy:2012va}
Benoit-Levy A., Smith K.~M., Hu W., 2012, Phys.Rev., D86, 123008

\bibitem[{Bond, Jaffe \& Knox(1998)Bond, Jaffe, \& Knox}]{Bond:1998zw}
Bond J., Jaffe A.~H., Knox L., 1998, Phys.Rev., D57, 2117

\bibitem[{{Born} \& {Wolf}(1980)}]{born/wolf:1980}
{Born} M., {Wolf} E., 1980, {Principles of Optics Electromagnetic Theory of
  Propagation, Interference and Diffraction of Light}

\bibitem[{{Bowyer}, {Jaffe} \& {Novikov}(2011){Bowyer}, {Jaffe}, \&
  {Novikov}}]{bowyer/etal:2011}
{Bowyer} J., {Jaffe} A.~H., {Novikov} D.~I., 2011, {MasQU: Finite Differences
  on Masked Irregular Stokes Q,U Grids}. Astrophysics Source Code Library

\bibitem[{{Bunn}(2011)}]{bunn:2011}
{Bunn} E.~F., 2011, PRD, 83, 083003

\bibitem[{{Cao} \& {Fang}(2009)}]{cao/fang:2009}
{Cao} L., {Fang} L.-Z., 2009, Astrophys.J, 706, 1545

\bibitem[{Das, Hajian \& Spergel(2009)Das, Hajian, \& Spergel}]{Das/etal:2009}
Das S., Hajian A., Spergel D.~N., 2009, Phys. Rev. D, 79, 083008

\bibitem[{Das {et~al}\mbox{.}(2013)Das, Louis, Nolta, Addison, Battistelli,
  {et~al.}}]{Das:2013zf}
Das S., Louis T., Nolta M.~R., Addison G.~E., Battistelli E.~S., {et~al.}, 2013

\bibitem[{Das {et~al}\mbox{.}(2011)Das, Marriage, Ade, Aguirre, Amir,
  {et~al.}}]{Das:2010ga}
Das S., Marriage T.~A., Ade P.~A., Aguirre P., Amir M., {et~al.}, 2011,
  Astrophys.J., 729, 62

\bibitem[{{Grain}, {Tristram} \& {Stompor}(2009){Grain}, {Tristram}, \&
  {Stompor}}]{grain/etal:2009}
{Grain} J., {Tristram} M., {Stompor} R., 2009, PRD, 79, 123515

\bibitem[{{Grain}, {Tristram} \& {Stompor}(2012){Grain}, {Tristram}, \&
  {Stompor}}]{grain/etal:2012}
{Grain} J., {Tristram} M., {Stompor} R., 2012, PRD, 86, 076005

\bibitem[{Hinshaw {et~al}\mbox{.}(2003)Hinshaw {et~al.}}]{Hinshaw:2003ex}
Hinshaw G., {et~al.}, 2003, Astrophys.J.Suppl., 148, 135

\bibitem[{Kamionkowski, Kosowsky \& Stebbins(1997)Kamionkowski, Kosowsky, \&
  Stebbins}]{Kamionkowski:1996ks}
Kamionkowski M., Kosowsky A., Stebbins A., 1997, Phys.Rev., D55, 7368

\bibitem[{Kermish {et~al}\mbox{.}(2012)Kermish, Ade, Anthony, Arnold, Arnold,
  {et~al.}}]{Kermish:2012eh}
Kermish Z., Ade P., Anthony A., Arnold K., Arnold K., {et~al.}, 2012

\bibitem[{Lewis(2005)}]{Lewis:2005tp}
Lewis A., 2005, Phys.Rev., D71, 083008

\bibitem[{{Lewis}, {Challinor} \& {Turok}(2002){Lewis}, {Challinor}, \&
  {Turok}}]{lewis/etal:2002}
{Lewis} A., {Challinor} A., {Turok} N., 2002, PRD, 65, 023505

\bibitem[{Niemack {et~al}\mbox{.}(2010)Niemack, Ade, Aguirre, Barrientos,
  Beall, {et~al.}}]{Niemack:2010wz}
Niemack M., Ade P., Aguirre J., Barrientos F., Beall J., {et~al.}, 2010,
  Proc.SPIE Int.Soc.Opt.Eng., 7741, 77411S

\bibitem[{Seljak \& Zaldarriaga(1997)}]{Seljak:1996gy}
Seljak U., Zaldarriaga M., 1997, Phys.Rev.Lett., 78, 2054

\bibitem[{Smith(2006)}]{Smith:2005gi}
Smith K.~M., 2006, Phys.Rev., D74, 083002

\bibitem[{Smith \& Zaldarriaga(2007)}]{Smith:2006vq}
Smith K.~M., Zaldarriaga M., 2007, Phys.Rev., D76, 043001

\bibitem[{Story {et~al}\mbox{.}(2012)Story, Reichardt, Hou, Keisler, Aird,
  {et~al.}}]{Story:2012wx}
Story K., Reichardt C., Hou Z., Keisler R., Aird K., {et~al.}, 2012

\bibitem[{{Zhao} \& {Baskaran}(2010)}]{zhao/baskaran:2010}
{Zhao} W., {Baskaran} D., 2010, PRD, 82, 023001

\end{thebibliography}

\appendix 
\section{Errors}
Here we derive an analytic expression for the expected error bars on each of the cross-power spectrum, X, Y, W and Z stand for T, E and B. The variance is given by
\ba
 \Theta^{(X \times Y);( W \times Z)}_{bb}  &=& \langle (C_{b} ^{(X \times Y)}-  \langle C_{b} ^{(X \times Y)} \rangle) (C_{b} ^{(W \times Z)}-  \langle C_{b} ^{(W \times Z)} \rangle) \rangle  \nonumber \\
  &=& \frac{1}{N} \frac{1}{\nu_b^2} \sum_{i, j, k, l}^{n_{d}}  \sum_{\vec{\ell} \in b}   \sum_{ \vec{\ell'} \in b} \left( \ave{X^{*i}_{\vec{\ell}} Y^{j}_{\vec{\ell}} W^{*k}_{\vec{\ell}'} Z^{l}_{\vec{\ell}' }} - \ave{C_{b} ^{(iX \times jY)} } \ave{C_{b} ^{(kW \times lZ)} } \right) \times  (1-\delta_{ij} )(1-\delta_{kl} ).
 \ea
The Kronecker symbol removes the auto power spectra. $n_{d}$ represents the number of splits we are cross correlating and  $\nu_b$ the number of modes in the annuli b.  The general normalization is
\ba
N= \sum_{i, j, k, l}^{n_{d}}  (1-\delta_{ij} )(1-\delta_{kl} ) = \sum_{i, j, k, l}^{n_{d}} \left(1-\delta_{ij}-\delta_{kl}+\delta_{ij}\delta_{kl} \right)  =n_{d}^{4}-2 n_{d}^{3}+n_{d}^{2}.
\ea
Applying Wick's theorem,
\ba
 \Theta^{(X \times Y);( W \times Z)}_{bb}  =  \frac{1}{N} \frac{1}{\nu_b} \sum_{i, j, k, l}^{n_{d}} \left[  \ave{C_{b} ^{(iX \times kW)} } \ave{C_{b} ^{(jY \times lZ)}} +  \ave{C_{b} ^{(iX \times lZ)} } \ave{C_{b} ^{(jY \times kW}}\right]  \times  (1-\delta_{ij} )(1-\delta_{kl} ).
\ea
We can decompose the estimated power spectrum into signal and noise, such that
\ba
 \ave{C_{b} ^{(iX \times kW)} }=C_{b} ^{X \times W} + \delta_{ik}N_{b} ^{X \times W},
\ea
and then we can decompose $ \Theta^{(X \times Y);( W \times Z)}_{bb}$ in three terms
\ba
 \Theta^{(X \times Y);( W \times Z)}_{bb}&=&\frac{1}{\nu_b}\left( C_{b} ^{X \times W}C_{b} ^{Y \times Z} +C_{b} ^{X \times Z}C_{b} ^{Y \times W} \right)  \nonumber \\
 &+&  \frac{1}{N} \frac{1}{\nu_b} \sum_{i, j, k, l}^{n_{d}}  \left( C_{b} ^{X \times W} \delta_{jl}N_{b} ^{Y \times Z} +C_{b} ^{Y \times Z}  \delta_{ik}N_{b} ^{X \times W}+C_{b} ^{X \times Z} \delta_{jk}N_{b} ^{Y \times W}+C_{b} ^{Y \times W}  \delta_{il}N_{b} ^{X \times Z} \right)\times  (1-\delta_{ij} )(1-\delta_{kl} ) \nonumber \\
 &+& \frac{1}{N} \frac{1}{\nu_b} \sum_{i, j, k, l}^{n_{d}} \left( \delta_{ik}N_{b} ^{X \times W} \delta_{jl}N_{b} ^{Y \times Z} +\delta_{il}N_{b} ^{X \times Z}\delta_{jk}N_{b} ^{Y \times W}  \right) \times  (1-\delta_{ij} )(1-\delta_{kl} ).
\ea
Finally, using
\ba
\sum_{i, j, k, l}^{n_{d}} \delta_{jl} \left(1-\delta_{ij}-\delta_{kl}+\delta_{ij}\delta_{kl} \right)  = n_{d}^{3}-2n_{d}^{2}+n_{d} ,
\ea
and
\ba
\sum_{i, j, k, l}^{n_{d}}  \delta_{ik}\delta_{jl}  \left(1-\delta_{ij}-\delta_{kl}+\delta_{ij}\delta_{kl} \right)= n_{d}^{2}-n_{d},
\ea
the variance is given by
\ba
 \Theta^{(X \times Y);( W \times Z)}_{bb}&=&\frac{1}{\nu_b}\left( C_{b} ^{X \times W}C_{b} ^{Y \times Z} +C_{b} ^{X \times Z}C_{b} ^{Y \times W} \right)  \nonumber \\
 &+& \frac{1}{\nu_b n_{d}} \left( C_{b} ^{X \times W} N_{b} ^{Y \times Z} +C_{b} ^{Y \times Z}  N_{b} ^{X \times W}+C_{b} ^{X \times Z} N_{b} ^{Y \times W}+C_{b} ^{Y \times W} N_{b} ^{X \times Z} \right) \nonumber \\
 &+& \frac{1}{\nu_b n_{d}(n_{d}-1)}  \left( N_{b} ^{X \times W} N_{b} ^{Y \times Z} +N_{b} ^{X \times Z}N_{b} ^{Y \times W}  \right).
\ea

\end{document}